\documentclass[preprint,12pt]{elsarticle}

\usepackage{epsfig}
\usepackage{graphicx}
\usepackage{amssymb}
\usepackage{amsthm}
\usepackage[acronym,toc,shortcuts]{glossaries}
\usepackage{multirow}
\usepackage{caption}
\usepackage{subcaption}
\usepackage[lined,linesnumbered,noend,ruled]{algorithm2e}
\usepackage{color}
\usepackage[table,xcdraw]{xcolor}
\usepackage{soul}
\usepackage{url}
\usepackage[numbers]{natbib}
\usepackage{epstopdf}

\makeglossaries
\newacronym{3GPP}{3GPP}{The 3rd Generation Partnership Project}
\newacronym{5GPP}{5GPP}{The 5th Generation Partnership Project}
\newacronym{AAA}{AAA}{Authentication, Authorization and Accounting}
\newacronym{AP}{AP}{Access Point}
\newacronym{API}{API}{Application Programming Interface}
\newacronym{APN}{APN}{Access Point Name}
\newacronym{BE}{BE}{best-effort}
\newacronym{BET}{BET}{Blind Equal Throughput}
\newacronym{BS}{BS}{Base Station}
\newacronym{BTP}{BTP}{Backhaul Transport Provider}
\newacronym{CA}{CA}{carrier aggregation}
\newacronym{CAPEX}{CAPEX}{capital expenditure}
\newacronym{CELL-ID}{CELL-ID}{cell identification ID}
\newacronym{CoMP}{CoMP}{Coordinated Multipoint}
\newacronym{CPU}{CPU}{central processing unit}
\newacronym{CQI}{CQI}{channel quality indicator}
\newacronym{CS}{CS}{central scheduler}
\newacronym{CSI}{CSI}{channel state information}
\newacronym{D2D}{D2D}{Device-to-Device}
\newacronym{DiffServ}{DiffServ}{Differentiated Services}
\newacronym{eICIC}{eICIC}{enhanced inter-cell interference cancellation}
\newacronym{eNodeB}{eNodeB}{evolved Node-B}
\newacronym{EPC}{EPC}{Evolved Packet Core}
\newacronym{EPS}{EPS}{Evolved Packet System}
\newacronym{E-UTRAN}{E-UTRAN}{Evolved Universal Terrestrial Radio Access Network}
\newacronym{GGSN}{GGSN}{Gateway GPRS Support Node}
\newacronym{GPS}{GPS}{global positioning system}
\newacronym{GRA}{GRA}{Grey relational analysis}
\newacronym{GTP}{GTP}{GPRS Tunneling Protocol}
\newacronym{HDFS}{HDFS}{Hadoop Distributed File System}
\newacronym{HetNet}{HetNet}{heterogeneous network}
\newacronym{HiveQL}{HiveQL}{Hive Query language}
\newacronym{HSS}{HSS}{Home Subscriber Station}
\newacronym{HTTP}{HTTP}{Hypertext Transfer Protocol}
\newacronym{ICIC}{ICIC}{inter-cell interference coordination}
\newacronym{ICN}{ICN}{information-centric network}
\newacronym{IEEE}{IEEE}{Institute of Electrical and Electronics Engineers}
\newacronym{IMEI}{IMEI}{International Mobile Station Equipment Identity}
\newacronym{IMSI}{IMSI}{International Mobile Subscriber Identity}
\newacronym{IntServ}{IntServ}{Integrated Service}
\newacronym{IP}{IP}{Internet Protocol}
\newacronym{JSON}{JSON}{JavaScript Object Notation}
\newacronym{KPI}{KPI}{key performance indicator}
\newacronym{LAC}{LAC}{location area code}
\newacronym{LTE}{LTE}{Long Term Evolution}
\newacronym{MADM}{MADM}{Multiple Attribute Decision Making}
\newacronym{MEW}{MEW}{multiplicative exponent weighting}
\newacronym{MME}{MME}{Mobility Management Entity}
\newacronym{MMF}{MMF}{Max-Min Fairness}
\newacronym{MO}{MO}{Mobile Operator}
\newacronym{MSISDN}{MSISDN}{Mobile Station International Subscriber Directory Number}
\newacronym{MT}{MT}{Maximum Throughput}
\newacronym{MVNO}{MVNO}{Mobile Virtual Network Operator}
\newacronym{NFV}{NFV}{network function virtualization}
\newacronym{NoSQL}{NoSQL}{Not Only SQL}
\newacronym{OAM}{OAM}{Operation, Administration and Management}
\newacronym{OPEX}{OPEX}{operating expenditure}
\newacronym{OS}{OS}{operating system}
\newacronym{OTT}{OTT}{over-the-top}
\newacronym{PCRF}{PCRF}{Policy and Charging Rules Function}
\newacronym{PDN}{PDN}{packet data network}
\newacronym{PF}{PF}{Proportional Fair}
\newacronym{PGW}{P-GW}{Packet Data Gateway}
\newacronym{PHY}{PHY}{physical layer}
\newacronym{PPP}{PPP}{{P}oisson point process}
\newacronym{PSD}{PSD}{power spectral density}
\newacronym{QoE}{QoE}{quality-of-experience}
\newacronym{QoS}{QoS}{quality-of-service}
\newacronym{RAN}{RAN}{radio access network}
\newacronym{RG}{RG}{Rate Guarantee}
\newacronym{RR}{RR}{Round Robin}
\newacronym{RSSI}{RSSI}{Received Signal Strength Indicator}
\newacronym{SAC}{SAC}{service area code}
\newacronym{SAW}{SAW}{simple additive weighting}
\newacronym{SCN}{SCN}{small cell network}
\newacronym{SDN}{SDN}{Software-Defined Networking}
\newacronym{SGSN}{SGCN}{Serving GPRS Support Node}
\newacronym{SGW}{S-GW}{Serving Gateway}
\newacronym{SHARING}{SHARING}{Self-organized Heterogeneous Advanced RadIo Networks Generation}
\newacronym{SINR}{SINR}{signal-to-interference-plus-noise ratio}
\newacronym{SLA}{SLA}{Service Level Agreement}
\newacronym{SSID}{SSID}{Service Set Identification}
\newacronym{ST}{ST}{Standart Multi-User TOPSIS}
\newacronym{SVD}{SVD}{singular value decomposition}
\newacronym{TEID}{TEID}{tunnel endpoint identifier}
\newacronym{TOPSIS}{TOPSIS}{Total Order Preference By Similarity to the Ideal Solution}
\newacronym{TTI}{TTI}{transmission time interval}
\newacronym{UE}{UE}{user equipment}
\newacronym{UMTS}{UMTS}{Universal Mobile Telecommunications Service} 
\newacronym{VoIP}{VoIP}{voice-over-IP}
\newacronym{WiFi}{WiFi}{Wireless Fidelity}
\newacronym{WLAN}{WLAN}{Wireless Local Area Network}
\newacronym{WMC}{WMC}{weighted Markov chain}
\newacronym{AI}{AI}{allocation interval}

\journal{Elsevier Computers \& Electrical Engineering Journal}

\begin{document}

\begin{frontmatter}

\title{Software-Defined Networking Based Network Virtualization for Mobile Operators}

\author{Omer~Narmanlioglu$^1$ and Engin~Zeydan$^2$}
\address{$^1$O. Narmanlioglu is with P.I. Works, Istanbul, Turkey 34912.\\
E-mail: {omer.narmanlioglu@piworks.net.}\\
$^2$E. Zeydan is with T\"{u}rk Telekom Labs, Istanbul, Turkey 34770.\\
E-mail: engin.zeydan@turktelekom.com.tr.}

\begin{abstract}

Software-Defined Networking (SDN) paradigm provides many features including hardware abstraction, programmable networking and centralized policy control. One of the main benefits used along with these features is core/backhaul network virtualization which ensures sharing of mobile core and backhaul networks among Mobile Operators (MOs). In this paper, we propose a virtualized SDN-based Evolved Packet System (EPS) cellular network architecture including design of network virtualization controller. After virtualization of core/backhaul network elements, eNodeBs associated with MOs become a part of resource allocation problem for Backhaul Transport Providers (BTPs). We investigate the performance of our proposed architecture where eNodeBs are assigned to the MOs using quality-of-service (QoS)-aware and QoS-unaware scheduling algorithms under the consideration of time-varying numbers and locations of user equipments (UEs) through Monte-Carlo simulations. The performances are compared with traditional EPS in Long Term Evolution (LTE) architecture and the results reveal that our proposed architecture outperforms the traditional cellular network architecture.

\end{abstract}


\end{frontmatter}

\newpage

\section{Introduction}

Developing new innovative solutions within current network infrastructure with respect to today's requirements is becoming difficult every day due to high complexity of networks~\cite{6461195}. It should be noted that even though the mobile technology is advancing rapidly, data transmission has been performed through the same backhaul since second generation of mobile technologies which is still valid in current \ac{LTE} systems. With the advancements in \ac{LTE}-Advanced and small cell technologies, backhaul of \glspl{MO} is expected to be similar to data centers with mesh network topologies. Taking into account these facts, the currently deployed network architecture of \ac{EPS} used in \ac{LTE} presents several drawbacks. Constantly deploying the infrastructure network equipments at high capacity is both costly and inefficient for \glspl{MO}. With respect to this, virtualizing the currently on demand network infrastructure owned by infrastructure owners, \glspl{BTP}, is crucial. Therefore, managing the dynamic network traffics resulting from users of \glspl{MO} and handling the possible complex \glspl{SLA} between \glspl{MO} and \glspl{BTP} via dynamic slicing becomes more important. However, current infrastructure deployment solutions cannot enable such virtualization option for \glspl{MO} due to lack of proper usage of recent technological advancements in network virtualization. Therefore, \glspl{MO} and \glspl{BTP} are looking for new innovative solutions in order to overcome the increasing demands of these network dynamics~\cite{6553676}.

Recent developments such as \ac{SDN}, which is initially implemented using OpenFlow protocols, provides powerful and simple approaches to manage complex networks by creating programmable, dynamic and flexible architecture, abstraction from hardware and centralized controller structure. In addition to \ac{SDN}, network virtualization has become one of the major recent innovations that can also provide flexible and scalable logical infrastructure to every organization. In respect to this, network virtualization with \ac{SDN} is an important paradigm that ensures the efficient usage of network resources. It can provide several features such as sharing of resources by breaking down the larger ones into multiple virtualized pieces, isolation of resources for better monitoring of data privacy and interference-free network access among users, aggregation for combining smaller resources into a single virtual resource, dynamism for fast deployment and reliable scalability in order to deal with the users' mobility, ease in resource management for debugging, testing and rapid deployment purposes~\cite{6658648}. On the other hand, developing appropriate scheduling mechanisms for resource allocation plays a fundamental role and help to meet \ac{QoS} requirements of applications used by \glspl{MO}' users. Depending on the application types (\ac{VoIP}, video conferencing, streaming media etc.), the requirements differ, however, they can be mapped to common parameters such as minimum guaranteed data rate, transmission delay, jitter and packet loss rate. Therefore, combining the advantages of \ac{SDN} for network virtualization with appropriate scheduling algorithms under the consideration of those \ac{QoS} parameters for dynamic resource allocations plays a key role in satisfying the demands of users associated with \glspl{MO} as well as of \glspl{BTP}.

\subsection{Related Works}

Before the introduction of \ac{SDN}, traditional \ac{QoS} providing approaches such as \glspl{IntServ}~\cite{ietf1994rfc} and \glspl{DiffServ}~\cite{baker1998definition}, had been standardized. However, there have been several drawbacks of these approaches. \ac{IntServ}, a fine-grained traffic control architecture, is only applicable for small scale networks. \ac{DiffServ}, on the other hand, is coarse-grained and applicable for larger networks, but it can only provide predefined/static 64 different classes of traffic to be differentiated since \ac{DiffServ} routers forward the packets based on $8-$bit DS field in the \ac{IP} header~\cite{baker1998definition}. This makes it hard for \ac{DiffServ} to fine tune the \ac{QoS} of separate flows. For example, the limit of DS field can be reached when four tenants each with sixteen application traffic types exists in the system. On the other hand, \ac{SDN} can enable fine-grained tuning (e.g. rules defined per flow) based on the specific application or user needs without restrictions. Therefore, approaches utilizing  more scalable techniques such as \ac{SDN} can provide better \ac{QoS} guarantees for big networks that have large coverage areas as in the case for \glspl{MO}.

New cellular network architectures based on \ac{SDN} paradigm have also been extensively investigated in the literature~\cite{li2012toward, jin2013softcell, pentikousis2013mobileflow, naudts2012techno, basta2014applying, tomovic2014sdn, gudipati2013softran}.  In~\cite{li2012toward}, \ac{SDN} architecture with four extensions to controller platforms, switches and base stations is proposed for cellular networks. The proposed \ac{SDN} architecture helps to simplify the design and management in order to address major limitations of today's cellular network architectures. A novel architecture, namely SoftCell, supporting fine-grained policies for cellular core network is proposed in~\cite{jin2013softcell} with the usage of packet classification on access switches that are next to the base stations and aggregation of traffic along multiple dimensions. In~\cite{pentikousis2013mobileflow}, software-defined based mobile network architecture that increases the operator's innovation potential is presented and validated by testbed implementation. \cite{naudts2012techno}~provides techno-economic analyses of two network scenarios which are software-defined non-shared and virtualized shared networks as well as comparisons with traditional networks. In~\cite{basta2014applying}, the authors point out applications of \ac{NFV} and \ac{SDN} while minimizing the transport network load overhead against several parameters (i.e., delays, number and placement of data centers etc.) as the function placement problem and aim to model and provide a solution for \ac{LTE} mobile core gateways. \cite{tomovic2014sdn}~examines several implementation scenarios of \ac{SDN} in mobile cellular networks and \ac{SDN}'s contributions to inter-cell interference management, traffic control and network virtualization domains are explained. SoftRAN~\cite{gudipati2013softran} abstracts all base stations in a local area as a virtual big base station that is managed by a centralized controller to perform load balancing, resource allocation, handover etc. while considering global view of the network. Moreover, although network sharing in the context of relationship between third parties (e.g. \glspl{MVNO}, vertical players) and \glspl{MO} has been widely discussed, most of the related works are in the context of economic advantages, business requirements and operational benefits that network sharing can introduce~\cite{6035827, frisanco2008infrastructure, 7054720, 7514161}. Recently, \ac{5GPP} has proposed some vertical use cases when envisioned 5G architecture (which can be owned by different entities such as mobile, transport or cloud infrastructure providers) provides infrastructure slices over the same physical infrastructure~\cite{5GPPref}. None of these approaches, however, consider the opportunity of applying both virtualization of mobile core/backhaul and as a consequence dynamic assignment of available \glspl{eNodeB} to different \glspl{MO} based on their traffic demands which come basically from their respective \glspl{UE} using various scheduling algorithms. This can be especially achieved by designing a \ac{SDN}-based shared \ac{EPS} architecture for multiple \glspl{MO} with a virtualization controller that is subject to instructions from \ac{BTP}.   

Many vendors such as Cisco, VMware, Big Switch, NEC etc. provide different approaches to network virtualization which are critical for infrastructure providers~\cite{instance1290}. Additionally, different network virtualization technologies have been studied in the literature~\cite{7295561, al2014openvirtex, sherwood2009flowvisor, 6385043}. In \cite{7295561}, the authors have classified different network hypervisors based on centralized and distributed architectures at the top level classification criterion and a second-level classification is executed based on the hypervisor execution platform. Technologies including OpenVirteX~\cite{al2014openvirtex} and FlowVisor~\cite{sherwood2009flowvisor} act as the transparent proxy between multiple controllers and forwarding elements that can create multiple slices of network resource  based on different slicing dimensions such as bandwidth, topology, forwarding table or device \ac{CPU}. FlowVisor is one of the enforcers of \ac{SDN} based network virtualization leveraging the capabilities of OpenFlow in order to provide network isolation between different slices. OpenVirteX (OVX) is another network virtualization platform developed by Stanford University's ON.LAB. OVX provides a different perspective and approach to FlowVisor. It also provides alternatives to virtual addressing strategies in order to keep spacing and separation among tenants, to virtual network topology so that tenants can be able to define their respective architectures while ensuring resilience for underlay networks. For supporting additional failover capabilities in case of congestion and failures within the network, VeRTIGO, which is an extension of FlowVisor, has been proposed in~\cite{6385043}. Even though all those approaches have different capabilities and present different performance-complexity trade-offs, generally using any of those approaches as a virtualization controller is able to meet the virtualization requirements of our proposed \ac{SDN}-based shared \ac{EPS} architecture.

\subsection{Our Contributions}

In this paper, we develop an \ac{SDN}-based virtualization controller architecture through a systematic modeling of virtual core and backhaul elements based on \ac{SDN} paradigm. In our developed scenario, a network virtualization controller, which is owned by \ac{BTP}, is directly connected to \ac{SDN} controllers of \glspl{MO}. After achieving the virtualization of core/backhaul network equipments, all \glspl{eNodeB} associated with different \glspl{MO} become a part of resource allocation problem for \glspl{BTP}. This virtualization controller connected to backhaul and core network elements is used to adaptively perform \ac{eNodeB} assignment to each \ac{MO} under the consideration of time-varying numbers and locations of associated \glspl{UE} and \glspl{MO}' demands. Using the proposed architecture, problem of \ac{eNodeB} assignment from an \ac{eNodeB} pool to each \ac{MO} is investigated. It should be noted that complex \glspl{SLA} between \glspl{MO} and \ac{BTP} can yield the necessity of solving multi-objective optimization problem. In order to approach this complex problem, we simply focus on solving one or two optimization parameters such as fairness, data rate and satisfied-MO-ratio. This optimization is performed with the use of \ac{QoS}-aware schedulers including \ac{MMF} and \ac{RG} and \ac{QoS}-unaware schedulers including \ac{RR}, \ac{BET}, \ac{MT} and \ac{PF} that are executed in virtualization controller. Moreover, their performances are compared with existing traditional \ac{EPS} architecture in \ac{LTE} through Monte-Carlo simulations. The results reveal that our proposed architecture outperforms the traditional \ac{EPS} in \ac{LTE} architecture depending on the selected scheduling method by \ac{BTP} with respect to considered optimization parameter(s). Additionally, we try to generate our proposed architecture with the use of \ac{MMF} scheduler on \textit{de-facto} \ac{SDN} emulator, namely Mininet~\cite{lantz2010network}, to showcase the benefits of applying scheduling algorithms to provide good service quality for \ac{QoS}-guaranteed \glspl{MO}. Our contributions in this paper can be summarized as follows: 
\begin{itemize}
  \item An \ac{SDN}-based shared EPS architecture is proposed in order to bring substantial benefits to \glspl{MO} and \glspl{BTP} which are responsible for setting up and maintaining \ac{EPC}, backhaul and \ac{E-UTRAN} of \ac{LTE} cellular systems.
  \item The performance of the proposed  \ac{SDN}-based shared \ac{EPS} architecture is investigated in terms of fairness, data rate and satisfied-\ac{MO}-ratio with the use of several \ac{QoS}-aware and \ac{QoS}-unaware schedulers. The performance improvements with respect to traditional \ac{EPS} used in \ac{LTE} cellular network architecture are demonstrated under the consideration of macro cell channel models with time-varying demands of \glspl{UE} associated with different \glspl{MO}.
\end{itemize}

The rest of the paper is organized as follows. In Section~II, we present our proposed virtualized \ac{EPS} architecture after describing the traditional \ac{EPS} in \ac{LTE}. In Section~III, the analytical expressions of scheduling algorithms for \ac{eNodeB} assignment to different \glspl{MO} are given and the performance of proposed architecture is evaluated in Section~IV. Finally, we conclude the paper in Section~V.

\section{System Model and Architecture}

\subsection{{Traditional \ac{EPS} architecture in \ac{LTE} networks}}

\begin{figure}[ht]
\centering
\includegraphics[trim={0 770 0 0},clip,width=\linewidth]{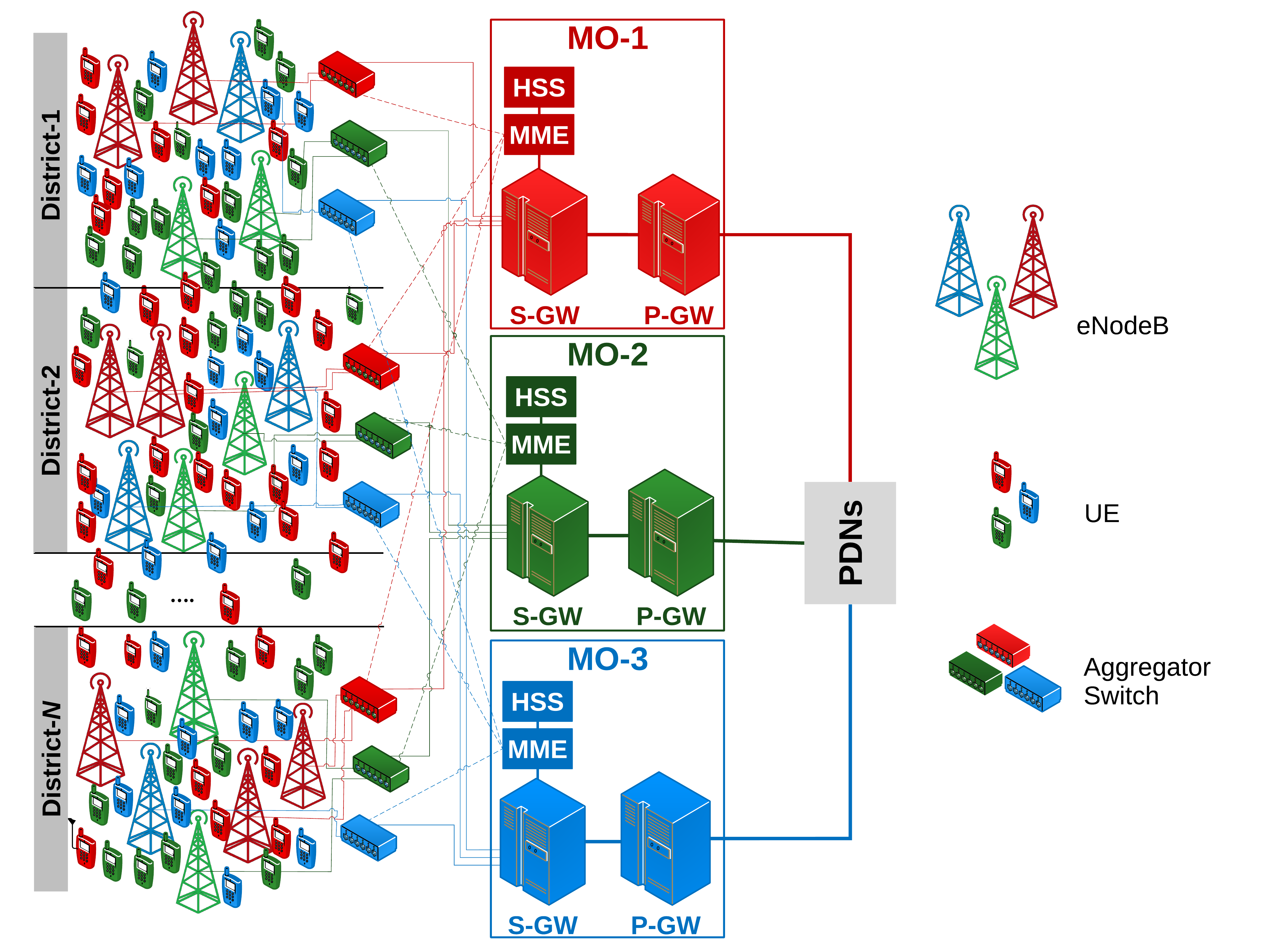}
\centering
\caption{Traditional \ac{EPS} architecture in \ac{LTE} networks with three \glspl{MO}.}
\centering
\label{trad}
\end{figure}

\ac{3GPP} has proposed \ac{EPS} architecture which is all-\ac{IP} based mobile network topology with less hierarchy and fewer nodes. \ac{EPC} and \ac{E-UTRAN} are the two main components of \ac{EPS}. Base station and mobile terminal in \ac{LTE} networks are denoted as \ac{eNodeB} and \ac{UE}, respectively. \ac{EPC} includes: (i) \ac{MME} in which user mobility, tracking, paging, roaming and intra-LTE handover are performed, (ii) \ac{SGW} that is responsible for routing and forwarding of packets among nodes and handover among \ac{LTE} and other cellular networks, (iii) \ac{PGW} that maintains the connection of LTE networks with the external \ac{IP}-based networks, (iv) \ac{HSS} that provides user identification and addressing. On the other hand, \ac{E-UTRAN} includes \glspl{eNodeB} that handle the communication between \glspl{UE} and core networks.

Fig.~\ref{trad} shows an example of a traditional \ac{EPS} architectural diagram of \ac{LTE} networks for three \glspl{MO}.  These \glspl{MO} are identified by different colors and they operate at the same geographical area where each \ac{MO} has its own core, backhaul and radio access network elements or has rented some of network equipments including mobile backhaul switches, routers from the main \ac{BTP} (as in some countries). The network equipments of each \ac{MO} (i.e., \glspl{eNodeB}, aggregator switches, \glspl{SGW} and \glspl{PGW}) are also identified by the same colors and all of them are connected to some \glspl{PDN} such as Internet through their own infrastructure.  It should be noted that in this architecture, none of the \glspl{MO} are sharing any resource or equipment and each of them has deployed its own network equipments independent of each other.

The locations and numbers of \glspl{eNodeB} as well as the capacity values of core/backhaul network elements associated with each \ac{MO} are predetermined. This is done under the consideration of statistics of network characteristics such as average \ac{UE} distributions, traffic loads, connection requests, etc. Thereby, they cannot be applicable for dynamic and adaptive changing network conditions which is one of the main drawbacks of this traditional \ac{EPS} in \ac{LTE} architecture. Moreover, due to the lack of dynamicity in the network, \glspl{MO} may operate in over or under capacity conditions in some situations. As a consequence, this can introduce inefficient utilisation of network capacity, higher \ac{CAPEX}/\ac{OPEX} and connection problems during disastrous events.

\subsection{Our Proposed Virtualized \ac{EPS} Architecture}

\ac{SDN} allows the capability of virtualization based on different scenarios including topology, hardware, device \ac{CPU} and bandwidth of individual links with priority settings within the network amongst \glspl{MO}. In this section, we are introducing a new shared \ac{EPS} architecture based on the \ac{SDN} concept for mobile core/backhaul virtualization.

\subsubsection{Virtualization Controller for Mobile Core and Backhaul Sharing}

The network virtualization can readily apply to the provisioning of a \ac{SDN}-based shared \ac{EPS} network which is owned by \ac{BTP} and utilized by \glspl{MO}. The streams of different \glspl{MO} are isolated from one another and each \ac{MO} can control its own allocated slice of the network without any regard to the other \glspl{MO} sharing the same network. The network slices allocated to individual \glspl{MO} can be managed by the \ac{BTP} via a virtualization controller (e.g., via a controller similar to FlowVisor~\cite{sherwood2009flowvisor}).

The \ac{SDN} framework also allows the \ac{BTP} to act as a broker in this setting to modify and adapt the slices in real time based on the agreements between the \ac{BTP} and the \glspl{MO}. The individual \glspl{MO} can then control their own slices via their dedicated control plane architectures (i.e., via their own \ac{MME}, \ac{HSS} and \ac{PCRF}). Every time a new rule needs to be pushed by an \ac{MO}'s controller, the virtualization controller first checks the integrity and validity of the rule and then forwards the rule to the corresponding forwarders in the network. The \ac{SDN} framework with the virtualization controller allows all nodes, including the network forwarding hardware and network gateways (\glspl{SGW} and \glspl{PGW}), \glspl{PDN} and backhaul networks to be shared by \glspl{MO}. It also provides granularity in what is shared in the network. In a shared network, the \glspl{MO} may maintain their own \glspl{eNodeB}, gateway elements, and \glspl{PDN} or they may also share some of the gateway elements and \glspl{PDN}. However, all \glspl{MO} participating in the shared network have to maintain their own control planes (\ac{MME}, \ac{PCRF} and \ac{HSS}) and those are used to control the network slices that are allocated by the virtualization controller which is managed by the \ac{BTP}. 

\begin{figure}[h]
\centering
\includegraphics[trim={0 0 0 0},clip,width=\linewidth]{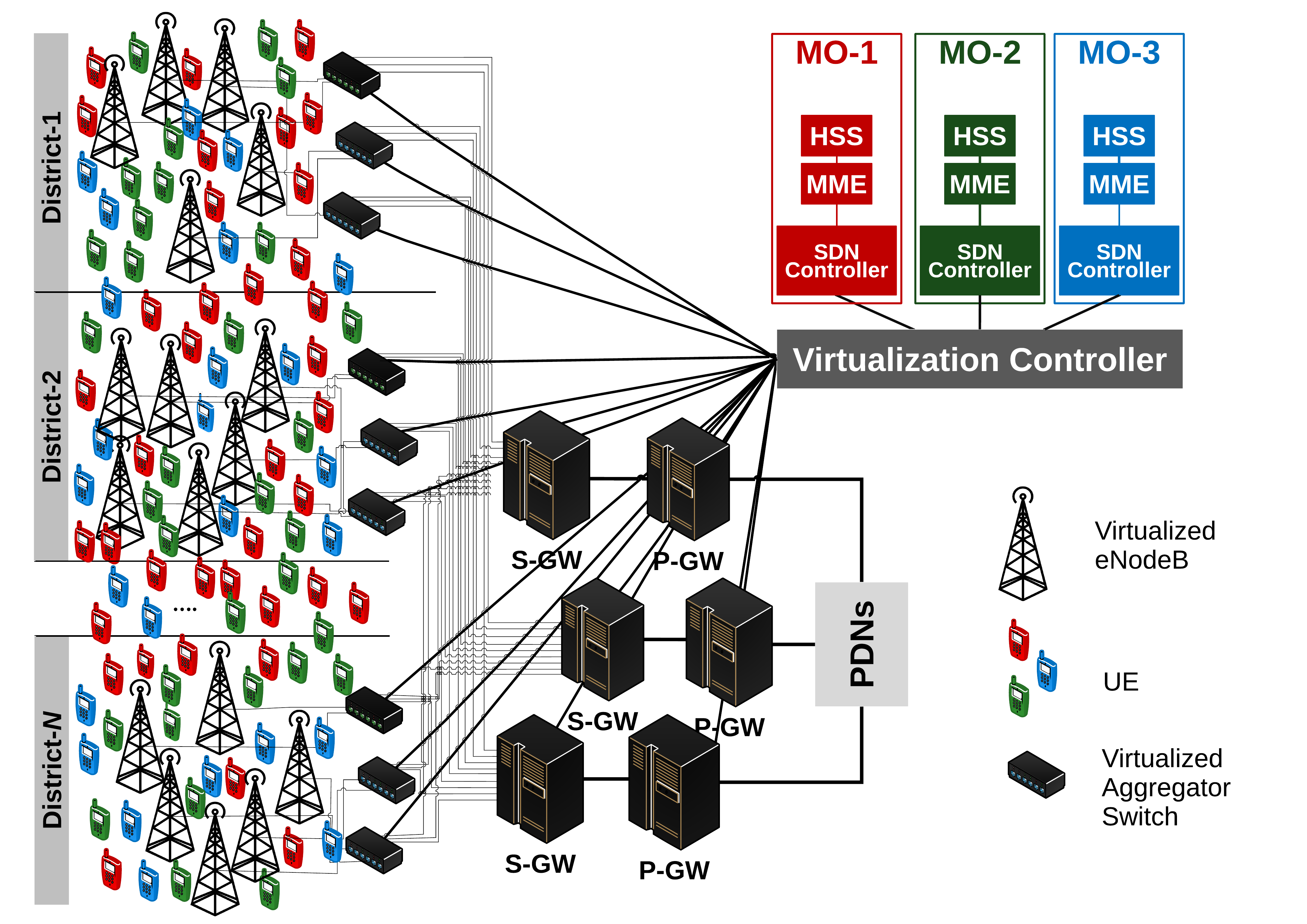}
\centering
\caption{The shared \ac{SDN}-based \ac{EPS} architecture in \ac{LTE} networks with three \glspl{MO}.}
\centering
\label{fig:Overview of SDN architecture}
\end{figure}

In our scenario (see Fig.~\ref{fig:Overview of SDN architecture}), a network virtualization controller that is directly connected to the \ac{SDN} controllers of each \ac{MO} is used to adaptively perform core/backhaul sharing between different \glspl{MO}. Note that \ac{SDN} controllers inside \glspl{MO} control the demand requests of each user as well as establish bi-directional communications with virtualization controller. Since each \ac{eNodeB} is connected to shared mobile backhaul and core gateways, such as \ac{SGW} via \ac{SDN}-based aggregator switches, sharing of mobile backhaul/core network equipments ensures availability of pools of \glspl{eNodeB}. The demands of multiple \glspl{MO} are served under time and location based varying network conditions by pushing appropriate flow rules on both backhaul and core network elements via protocols such as OpenFlow. The districts of Figs.~\ref{trad}~and~\ref{fig:Overview of SDN architecture} demonstrate the situations where a pool of \glspl{eNodeB} is available for multiple \glspl{MO} in both traditional and proposed architecture. In the districts of Fig.~\ref{trad}, traditional \glspl{eNodeB} assignment is depicted. When virtualization of core/backhaul network equipments is performed for each \ac{MO}, all \glspl{eNodeB} turn into a pool of radio access equipments owned by \ac{BTP} (as in districts of Fig.~\ref{fig:Overview of SDN architecture}). These radio access equipments can also be dynamically assigned to different \glspl{MO} with respect to their number of active \glspl{UE}, their varying demands and locations. This can provide many advantages such as efficient usage of network devices (e.g. availability of pool of \glspl{eNodeB}), balancing traffic demand/usage behaviour via dynamic scaling of the network, automation of provisioning services and multi-tenancy for multiple \glspl{MO}. In the shared mobile infrastructure, \ac{BTP}, the infrastructure owner allocates the slices including \glspl{eNodeB} and core/backhaul network equipments to \glspl{MO} based on the assignment decision of virtualized \glspl{eNodeB}. After that, all other \ac{RAN} related \ac{eNodeB} control functionalities including handover, roaming and radio resource management, \ac{CA} as well as interference management via \ac{CoMP} transmission and reception, \ac{eICIC} etc. are managed by the virtualization controller of \ac{BTP}, which is connected to the virtualized \glspl{eNodeB} through the virtualized aggregator switches. The virtualization controller also communicates with associated \glspl{MO} for those functionalities since all \glspl{MO} still maintain their own control planes including \ac{MME}, \ac{PCRF} and \ac{HSS}.

In this architecture, virtualization is performed at two levels. First, \glspl{BTP} manage the network slices assigned to each \ac{MO} using network virtualization controller. Second, sub-virtualization for all \ac{MO}'s applications is performed within a mobile operator's slice. In this \ac{SDN}-based \ac{EPS} architecture, traffic of multiple \glspl{MO} is converged to run on a common backbone network infrastructure while each stream of each \ac{MO} is kept virtually separate. In \ac{E-UTRAN}, all \glspl{UE} are assumed to be under the coverage of multiple \glspl{eNodeB} whose combination is abbreviated as District$-i$, for $i=\{1,2,...,N\}$ (pool of \glspl{eNodeB}), in Fig.~\ref{fig:Overview of SDN architecture} and they can be assigned to different \glspl{MO}. Note that based on this scenario, sharing of mobile network backhaul equipments among multiple \glspl{MO} and dynamic assignments of each \ac{eNodeB} to different \glspl{MO} through scheduling algorithms result in lower \ac{CAPEX} and \ac{OPEX} for both \glspl{MO} and \glspl{BTP}.

\subsubsection{System Model and \ac{eNodeB} Assignment Mechanism}

A cellular network topology can be separated into several districts that do not overlap and do not interfere with each other. A single district is composed of a set $\xi$ of \glspl{eNodeB}, thereby, a set $\xi$ of connections to an aggregator switch that is connected to \ac{SGW} and a set $\zeta$ of \glspl{MO} which have different number of associated \glspl{UE} and their time-varying demands, denoted by \textbf{$\Omega$}, are considered. The performance parameters (i.e., data rate or capacity) give different values as a result of the connection and communication with \glspl{MO} due to the variation in the total number of active \glspl{UE} associated with these \glspl{MO} and some physical factors such as path attenuation and fading. These factors lead to different signal quality at receiver side and different capacity levels on the links between \glspl{eNodeB} and \ac{SGW} under the assumption that backhaul links are not bottleneck. 

All possible links between \glspl{eNodeB} and \glspl{MO} can be modelled as a matrix, including all outcomes of set $\zeta$ (in row) and $\xi$ (in column). When the matrix is denoted by $\rm{\bf{M}}=\mathit{(m_{i,k})}_{|\zeta|x|\xi|}$, the value of predefined performance parameter or metric with respect to the combination of the $i^{th}$ \ac{MO} and the $k^{th}$ \ac{eNodeB} is denoted by $m_{i,k}$. Each connection will be assigned to a specific \ac{MO} and isolated from other \glspl{MO} through a virtualization controller. Now, we define an assignment vector, $\bf{\Phi}$, whose length is the same with the set $\bf{\xi}$ and it includes the indices of \ac{MO} associated with the connections. For instance, the vector of $\bf{\Phi}  = \left[ {\begin{array}{*{20}{c}} 2&1&{...}&1 \end{array}} \right]$ shows that the first connection (or \ac{eNodeB}) is assigned to the second \ac{MO}. Similarly, the second and last connections are assigned to the first \ac{MO}. We are interested in finding the best assignment vector which provides the highest predefined performance parameters or metric(s) such as fairness, maximum data rate or any other parameter related to \ac{QoS} requirements.

In this work, without loss of generality, we integrate the virtualization controller, which is controlled by \ac{BTP} explained in the previous section, into a single district for a general network topology. The virtualization controller aims at finding an assignment vector, $\bf{\Phi}$, using the various scheduling algorithms in our proposed architecture. The next section investigates those scheduling algorithms that are used by \ac{BTP} throughout the proposed \ac{SDN}-based \ac{EPS} architecture.

\section{Schedulers for Dynamic \ac{eNodeB} Assignments in Shared \ac{EPS} Architecture}

Schedulers are the core component of resource management for optimization of the network performance. In this section,  we briefly introduce their concepts and analytical expressions. Then, we provide a method to integrate those scheduling algorithms into the \ac{eNodeB} assignment mechanism in our proposed architecture.

Basically, schedulers allocate the transmission resources to different users who have different \ac{QoS} requirements and signal quality according to their allocation algorithms. The running algorithms in schedulers improve the system performance under the consideration of some performance metrics (i.e., total data rate, fairness among users) or \ac{QoS} requirements. The algorithms have different input parameters, leading to different performances and system complexities. However, their allocation mechanisms are commonly based on giving the $k^{th}$ resource to the $i^{th}$ user, if its metric ($m_{i,k}$) satisfies the following equation, 
\begin{equation}
{m_{i,k}} = \arg {\max _j}\left\{ {{m_{j,k}}} \right\}.  
\end{equation}

In this paper, we classify the schedulers into three categories which are \textit{channel-unaware}, \textit{channel-aware /} \textit{\ac{QoS}-unaware} and \textit{channel-aware /} \textit{\ac{QoS}-aware}. Table~\ref{Classification of Schedulers} shows two schedulers per category.

\begin{table}[ht]
\centering
\caption{Classification of {considered} schedulers.}
\label{Classification of Schedulers}
\begin{tabular}{|c|c|c|}
\hline
                  &  \ac{QoS}-aware    & \ac{QoS}-unaware       \\ \hline
Channel-aware     &  \ac{MMF}, \ac{RG} & \ac{MT}, \ac{PF}       \\ \hline
Channel-unaware   &    $-$             & \ac{BET}, \ac{RR}      \\ \hline
\end{tabular}
\end{table}

\textit{Channel-unaware schedulers} do not consider signal quality and \ac{QoS} requirements. They use the simplest algorithms while allocating the resources to users. \ac{RR} and \ac{BET} are \textit{channel-unaware schedulers} which have fair allocation mechanisms. In \ac{RR}, resource allocation is performed in a cyclic order of users and provides the fairness in terms of the number of allocated resources. Similarly, \ac{BET} aims at satisfying the fairness in terms of data rate among users by using the past average data rate of each user. Its performance metric can be defined as inverse of $\lambda_{i}(t)$ where ${\lambda _i}\left( t \right)$ denotes the average data rate of the ${i^{th}}$ user and it can be calculated by
\begin{equation}
\lambda _i (t) = \left ( 1-\frac{1}{\tau} \right ) \times \lambda _i (t-\Delta t)+\sum_{k}^{ }\frac{\delta _{i,k}(t-\Delta t) \times R_{i,k}(t-\Delta t)}{\tau},
\end{equation}
\noindent
where $\tau >1$ denotes time constant of smooth filter, $\Delta t$ is the \ac{AI}, which is the period of allocation, and $R_{i,k}(t)$ is the instantaneous achievable data rate of the $i^{th}$ user with the use of $k^{th}$ resource, and $\delta _{i,k}(t) \in \{0,1\}$ takes the value of $1$ if $i^{th}$ user is allocated with $k^{th}$ resource at $t^{th}$ time, otherwise, it is $0$.

In contrast, allocation mechanism of \textit{channel-aware schedulers} depends on \ac{CSI} that identifies the signal quality at receiver side and achievable data rates are estimated based on the periodically provided \ac{CQI} values. \ac{MT} and \ac{PF} schedulers fall into this category. The target of \ac{MT} scheduler is to maximize the total data rate of the system by exploiting user diversity. Therefore, its metric can be written as
\begin{equation}
m_{i,k} = R_{i,k}(t).
\end{equation}
However, it should be noted that \ac{MT} is totally unfair since cell-edge users cannot be allocated with any resource due to bad-channel conditions. On the other hand, \ac{PF} scheduler benefits from user diversity gain and considers proportional fairness among users. Therefore, \ac{PF} partially satisfies both data rate and fairness. The allocation mechanism of \ac{PF} can be thought of as a combination of \ac{BET} and \ac{MT} {algorithms}. Past average data rate as weighting factor is added into \ac{MT} algorithm. The performance metric of \ac{PF} algorithm for the $i^{th}$ user can be written as
\begin{equation}
{m_{i,k}} = \frac{R_{i,k}(t)}{\lambda_{i}(t)}.
\end{equation}
For generalized form of \ac{PF} scheduler, it is defined as, 
\begin{equation}
{m_{i,k}} = \frac{R_{i,k}(t)^\alpha }{\lambda_{i}(t)^\gamma}.
\end{equation}
\noindent
where $\alpha$ and $\gamma$ are the weighting factors of data rate and fairness, respectively.

None of \ac{RR}, \ac{BET}, \ac{MT} and \ac{PF} algorithms take into account the \ac{QoS} requirements. In general, the performance metric for \ac{QoS} scheduler algorithms is written as~\cite{singh2007normalized}
\begin{equation}
m_{i,k} = R_{i,k}(t) \times U^{'}(\lambda_{i}(t)),
\end{equation}
where $U\left( {{\lambda_{i}(t)}} \right)$ is defined as utility functions and $(.)^{'}$ denotes the first order derivative operator. Utility function for QoS-aware users in \ac{RG} scheduler~\cite{singh2007normalized}, is designed as
\begin{equation}
U\left( \lambda _i (t) \right) =  \left ( \Omega _i \right ) \times \left( {\log \left( \lambda _i (t) \right) + 1 - {e^{\left( { - \beta _i \frac{{\lambda _i (t) - {\Omega _i}}}{{{\Omega _i}}}} \right)}}} \right) ,
\end{equation}
where $\beta _i$ is a positive constant used to control the aggressiveness depending on the ratio of allocated resources and demands and $\Omega_i$ denotes the time-varying demand of the $i^{th}$ user. 

Another \ac{QoS}-aware scheduler is \ac{MMF} algorithm~\cite{nace2008max}, in which resources are allocated to users orderly and with respect to their increasing demands (i.e., data rates) and unsatisfied users are equally allocated with the remaining resource. 

To summarize all the scheduling algorithms discussed above, the pseudo code that finds the best assignment vector for each scheduling algorithm is given in Algorithm~\ref{algo:score} where $\rm{\bf{R}}=\mathit{(R_{i,k})}_{|\zeta|x|\xi|}$ and $\rm{\bf{\Omega}}={(\Omega_{i})}_{|\zeta|x1}$ denote the maximum achievable data rate matrix and demand vector respectively. In our case, allocated resources are \glspl{eNodeB} and users are \glspl{MO}. All scheduling mechanisms can be implemented in virtualization controller using this algorithm.

\begin{algorithm}[h]
\SetAlgoNoEnd
\KwIn{$\bf{R}$, $\bf{\Omega}$}
\KwOut{$\bf{\Phi}$}
initialization: set $\bf{\Phi}$ to zero\;
calculate $\rm{\bf{M}}$\;
\ForEach{$e$ \textbf{in} $\xi$}{
    ${\bf{\Phi}} ($\textit{e}$) = \mathop {\arg \max }\limits_{\mathit{v}  \in \zeta } \left[ {{{\mathit{m}}_{\mathit{v} ,e}}} \right]$\;
}
return $\bf{\Phi}$\;
\caption{Scheduling algorithms running at virtualization controller that assigns \glspl{eNodeB} to each \ac{MO}.}
\label{algo:score}
\end{algorithm}

\section{Performance Evaluations}

In this section, we perform both simulation and emulation studies in order to investigate the performance of our proposed architecture with the use of several \ac{QoS}-aware and \ac{QoS}-unaware schedulers that are used for assignments of \glspl{eNodeB} to \glspl{MO}. In our proposed architecture, there exists a shared \ac{SDN}-based \ac{EPS}  infrastructure in \ac{LTE} networks with multiple \glspl{MO} sharing resources from a pool of \glspl{eNodeB}. Traditional cellular network architectures of \ac{EPS} in \ac{LTE} networks, where \glspl{eNodeB} are statically assigned to each \ac{MO}, are considered as benchmark.

\begin{figure}[ht]
\centering
\begin{subfigure}{.32\textwidth}
\centering
\includegraphics[trim= 0 100 0 5,clip,width=0.8\linewidth]{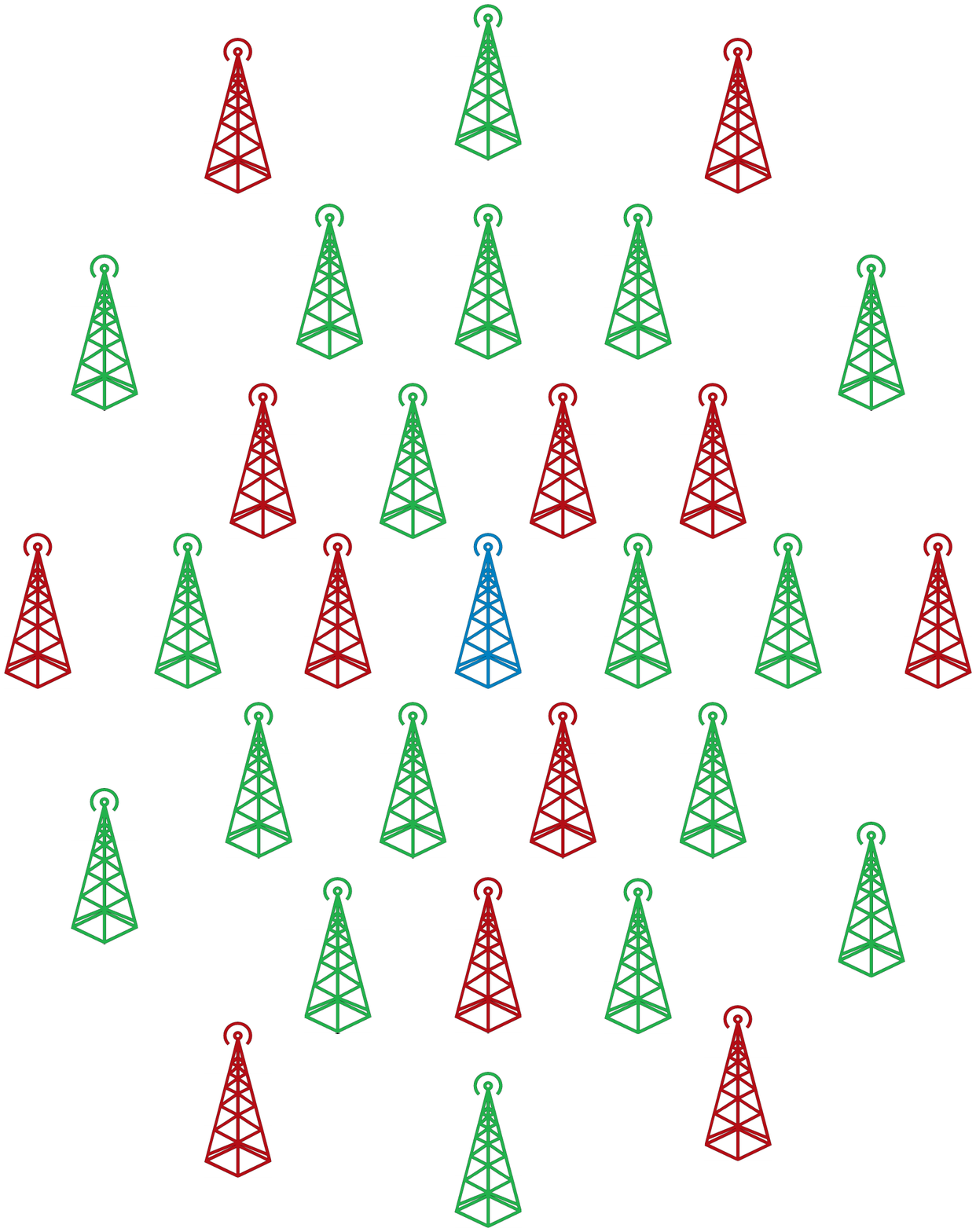}\\
(a)
\end{subfigure}
\begin{subfigure}{.32\textwidth}
\centering
\includegraphics[trim= 0 100 0 5,clip,width=0.8\linewidth]{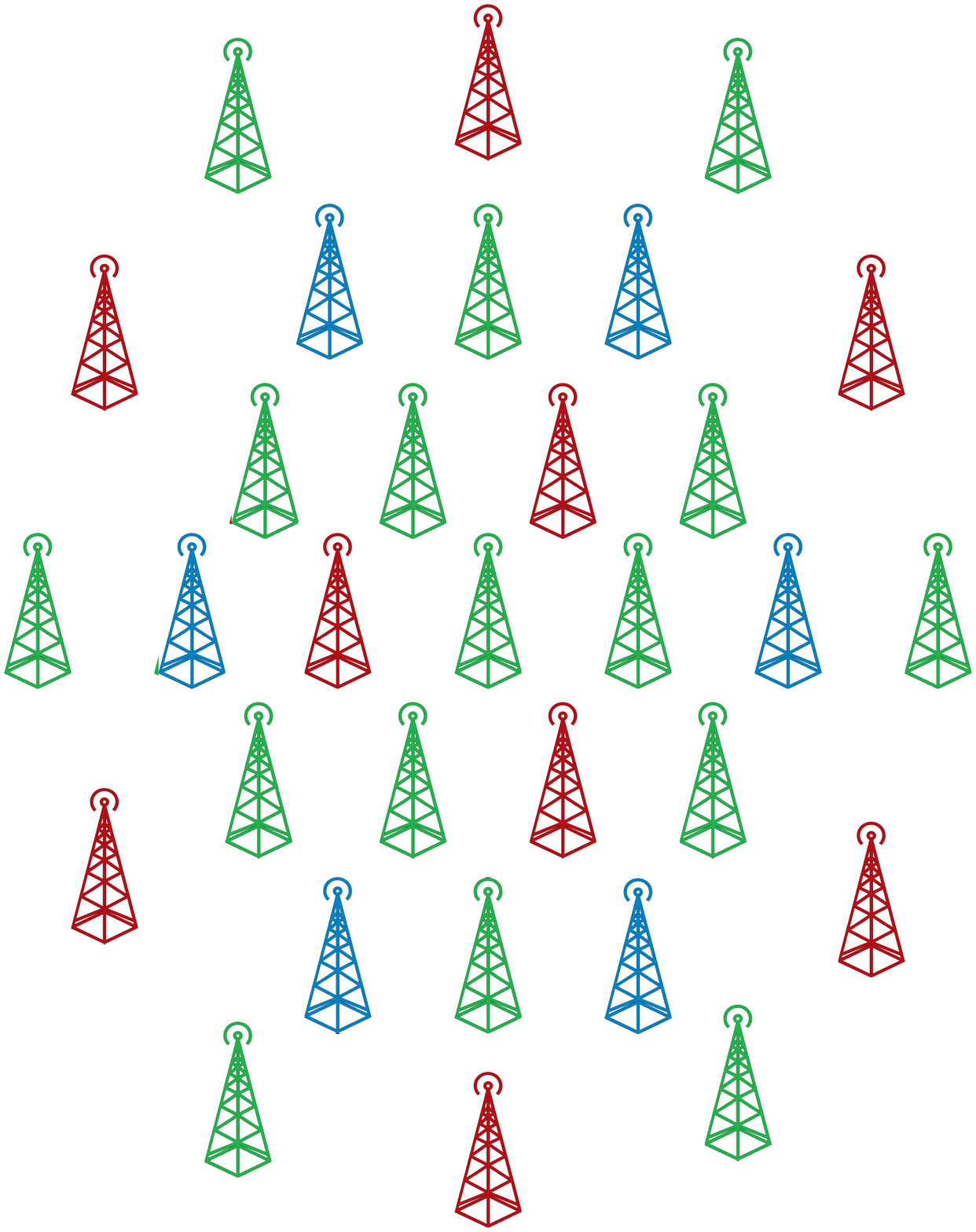}\\
(b)
\end{subfigure}
\centering
\begin{subfigure}{.32\textwidth}
\centering
\includegraphics[trim= 0 100 0 5,clip,width=0.8\linewidth]{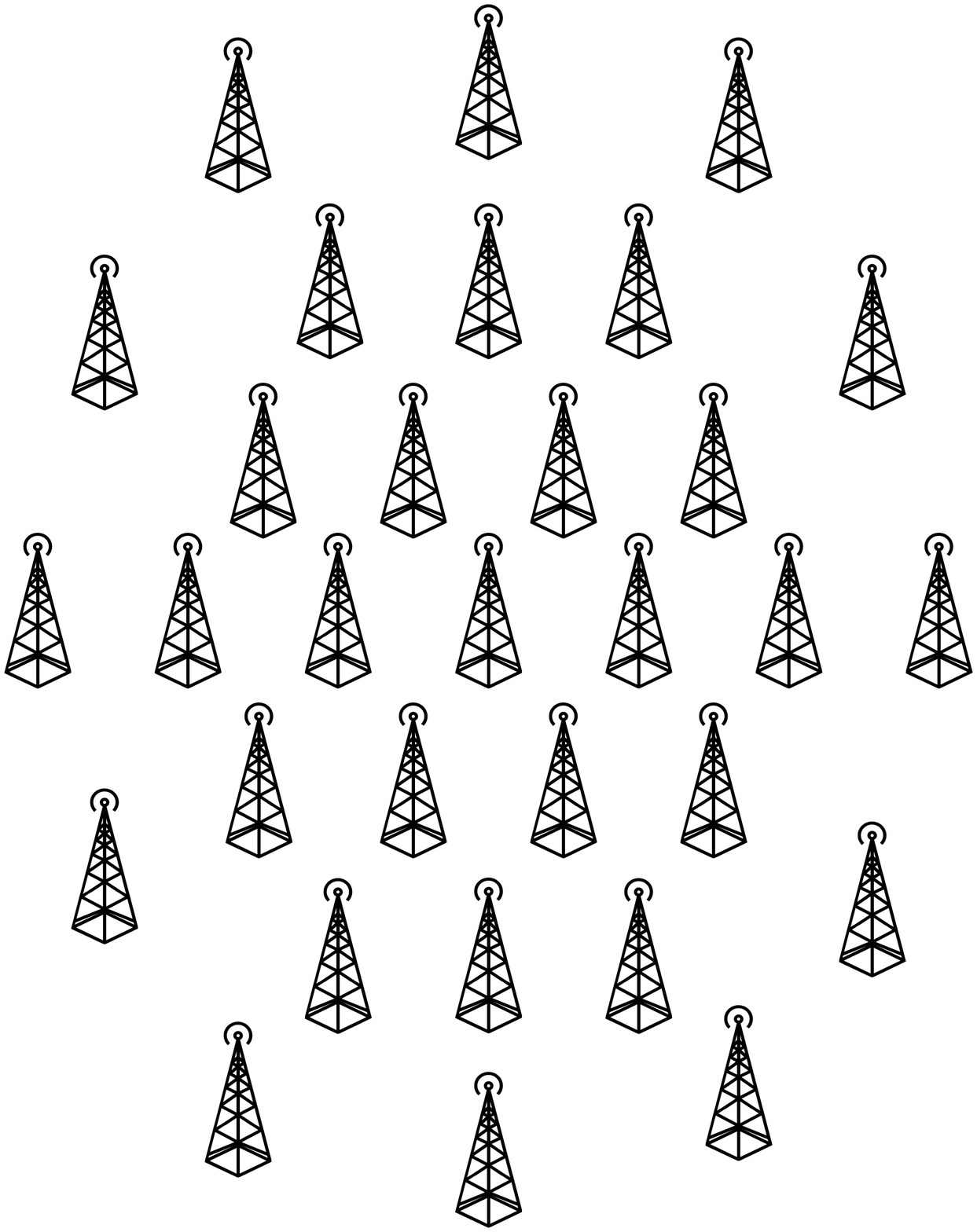}\\
(c)
\end{subfigure}
\centering
\caption{Locations of \glspl{eNodeB} when (a) \textit{demand-based} static assignment, (b) \textit{\ac{UE}-based} static assignment and (c) virtualized architecture are considered.}
\label{fig: Cell structure with multiple eNodeBs.}
\end{figure}

We assume that there exist $N$ non-overlapping districts (district$-1$, ... , district$-N$) connected to a single shared \ac{SGW} through aggregator switches. In our first simulation environment, \glspl{UE} are uniformly and \glspl{eNodeB} are deterministically distributed in the district$-1$ with the radius of $35$ $\rm{km}$, as shown in Fig.~\ref{fig: Cell structure with multiple eNodeBs.}. Relative simulation parameters are given in Table~\ref{Simulation parameters}. Demands of \ac{QoS}-aware \glspl{MO} which are \ac{MO}$-1$ and \ac{MO}$-2$ are uniformly distributed $(\rm{unif})$ and \ac{MO}$-3$ is considered as \ac{BE} operator. In \ac{RG} algorithm, $\beta$s for \ac{MO}$-1$ and \ac{MO}$-2$ are selected as $10$ and $9.5$, respectively (note that $\beta$ is zero for \ac{MO}$-3$). Additionally, time constant of smooth filter in \ac{BET}, \ac{PF} and \ac{RG} schedulers are set to $50$ and the generalized form of \ac{PF} with $\alpha$ of $1$ and $\gamma$ of $0.8$ is considered during simulations.

We consider two different static \ac{eNodeB} assignments for \glspl{MO} which are \textit{demand-based} and \textit{\ac{UE}-based} assignments in traditional \ac{EPS} architecture. In \textit{demand-based} assignment (see Fig.~\ref{fig: Cell structure with multiple eNodeBs.}~(a)), the numbers of \glspl{eNodeB} associated with \ac{MO}$-1$, \ac{MO}$-2$ and \ac{MO}$-3$ are set to $12$, $18$ and $1$ in order to keep the proportionality with their demands. On the other hand, in \textit{\ac{UE}-based} assignment (see Fig.~\ref{fig: Cell structure with multiple eNodeBs.}~(b)), the numbers of \glspl{eNodeB} associated with \ac{MO}$-1$, \ac{MO}$-2$ and \ac{MO}$-3$ are set to $9$, $16$ and $6$ so that they remain proportional to the numbers of their associated \glspl{UE} as in Table~\ref{Simulation parameters}. In both \textit{demand-based} and \textit{\ac{UE}-based} static assignments, \glspl{eNodeB} are homogeneously distributed in the considered hexagonal district structure. In addition, our proposed architecture with virtualized \glspl{eNodeB} is depicted in Fig.~\ref{fig: Cell structure with multiple eNodeBs.}~(c).

Under the consideration of urban and suburban areas in macro cell structure and the carrier frequency of $2$~$\rm{GHz}$, the channel gain, denoted by $H$, between \glspl{UE} and \glspl{eNodeB} depending on path loss and shadowing effects can be calculated by
\begin{equation}
H = 128 + 37.6 \mathrm{log}(d) + \psi  \quad [\mathrm{dB}],
\end{equation}
where $d$ is the distance between \ac{UE} and \ac{eNodeB} in $\rm{km}$ and $\psi$ (in $\rm{dB}$) is log-normal distributed (with zero mean and standard deviation of $8$) shadowing effect. While calculating the maximum achievable data rate, we use Shannon Capacity ($C$) which can be written as 
\begin{equation}
C = B \times {\log _2}\left( {1 + \frac{{PH}}{{{N_0}B}}} \right),
\end{equation}
where $B$, $P$ and $N_0$ are respectively bandwidth $\rm{(Hz)}$, transmit power $\rm{(W)}$ and noise \ac{PSD} $\rm{(W/Hz)}$ under the consideration of proper frequency spectrum sharing and advanced modulation techniques that ensure the interferences from neighbor \glspl{eNodeB} to be insignificant. In order to obtain stable and confident simulation results, $1000$ independent simulation runs are conducted. In each simulation, the locations of \glspl{UE} are randomly selected. In addition to this, each independent simulation has $1000$ time slots, in which shadowing effect and demands of users vary respectively in each and every fifty'th slots. 

\begin{table}[ht]
\centering
\caption{Simulation parameters.}
\label{Simulation parameters}
\begin{tabular}{|r|l|c|c|}
\hline
\multicolumn{1}{|c|}{}    & \multicolumn{1}{c|}{\ac{MO}$-1$}   & {\ac{MO}$-2$}    & {\ac{MO}$-3$}   \\ \hline
Number of \glspl{UE}      & \multicolumn{1}{c|}{$300$}         & { $500$}         & { $200$}        \\ \hline
Demand ($\rm{Gbps}$)      & \multicolumn{1}{c|}{$\rm{unif}(0,8)$} & { $\rm{unif}(0,12)$} & { $-$ }  \\ \hline
Carrier frequency         & \multicolumn{3}{l|}{$2$ $\rm{GHz}$}                                     \\ \hline
Bandwidth per user        & \multicolumn{3}{l|}{$5$ $\rm{MHz}$}                                     \\ \hline
Transmit power            & \multicolumn{3}{l|}{$46$ $\rm{dBm}$}                                    \\ \hline
Transmit power allocation & \multicolumn{3}{l|}{Uniform}                                            \\ \hline
Noise \ac{PSD}            & \multicolumn{3}{l|}{$-179$ $\rm{dBm/Hz}$}                               \\ \hline
\end{tabular}
\end{table}

\begin{figure}[ht]
\centering
\includegraphics[clip,trim=30 50 50 80,width=0.7\linewidth]{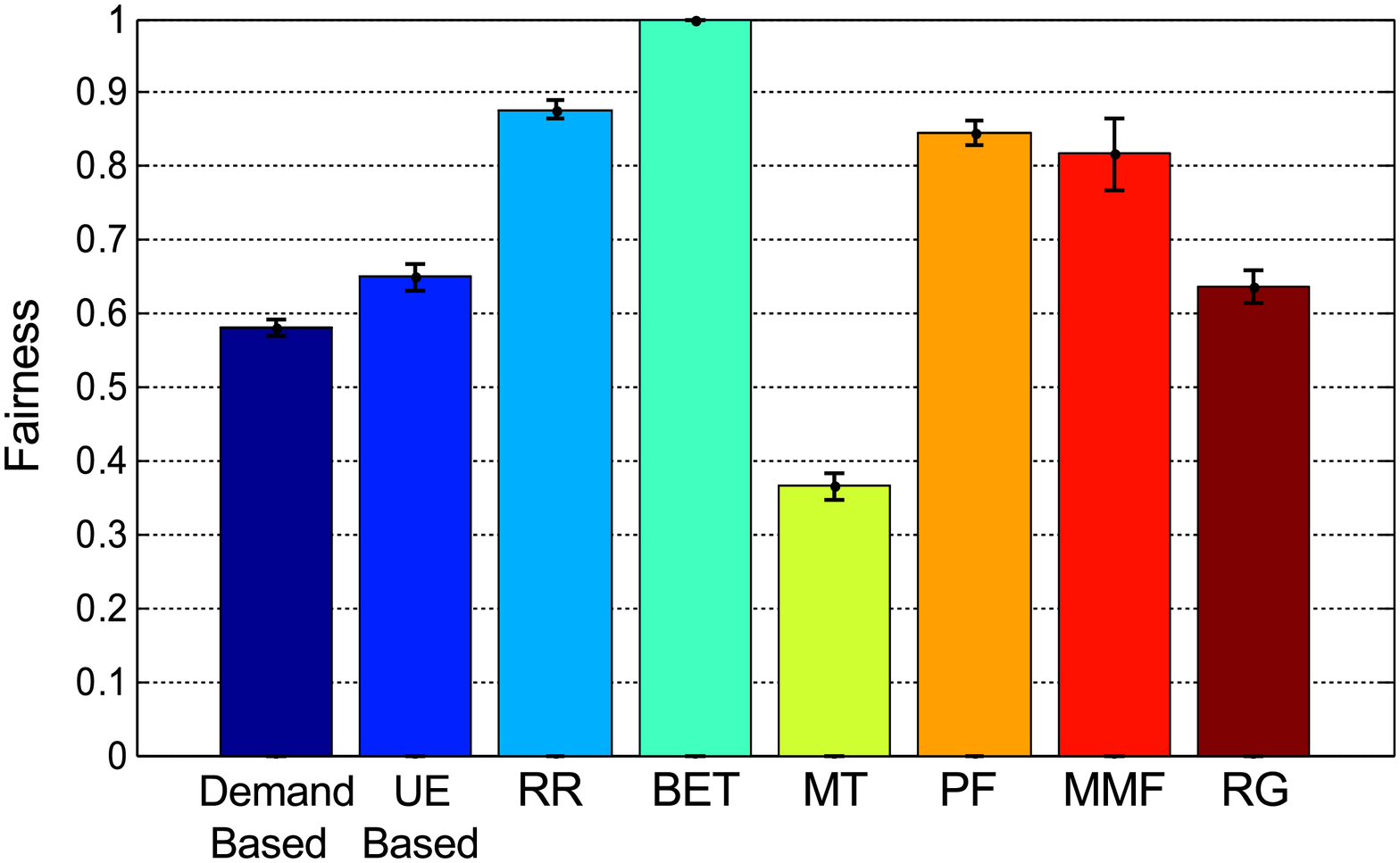}
\caption{Fairness performances of traditional \ac{EPS} in \ac{LTE} (\textit{demand-based} and \textit{\ac{UE}-based} \ac{eNodeB} assignments) and proposed shared \ac{SDN}-based \ac{EPS} architecture with different scheduler algorithms.}
\label{fig: Fairness indexes}
\end{figure}

In Fig.~\ref{fig: Fairness indexes}, we show the average and the standard deviations of fairness values for traditional \ac{EPS} in \ac{LTE} architecture based \ac{eNodeB} assignments with both \textit{demand-based} and \textit{\ac{UE}-based} and our proposed architecture with the use of \ac{RR}, \ac{BET}, \ac{MT}, \ac{PF}, \ac{MMF} and \ac{RG} schedulers as detailed in Algorithm~\ref{algo:score}. Jain's fairness index is used as a fairness metric which can be defined as
\begin{equation}
\textrm{Fairness}=  \frac{\left ( \sum_{i=1}^{M} R_{i} \right )^{2}}{M \times \sum_{i=1}^{M} R_{i}^2},
\end{equation}
where $M$ is the total number of \glspl{MO}.

Since \ac{BET} scheduler uses past average data rate as an inversely weighting factor during allocation, \ac{MO} with the highest data rate in current \ac{AI} has less chance of obtaining resource in the next \ac{AI} period. Therefore, \ac{BET} satisfies fairness among \glspl{MO} and presents the highest fairness index with the value of $0.99$. Relatively, the other fair allocation algorithm, \ac{RR} is the second with the highest value of $0.88$ since it allocates the sources in a cyclic order. The other fairness based schedulers which are \ac{PF} and \ac{MMF} give $0.85$ and $0.82$ fairness indexes, respectively. When we turn to our benchmarks, \textit{demand-based} assignment has $0.58$ and \textit{\ac{UE}-based} assignment has $0.65$ fairness values. The results show that our proposed architecture with \ac{RR}, \ac{BET}, \ac{PF} and \ac{MMF} outperforms both \textit{demand-based} and \textit{\ac{UE}-based} assignments and it improves the fairness index. The reason for this improvement is due to the fact that the metrics of all four schedulers consider fairness issue and adopt the allocation mechanism with respect to time-varying \ac{UE} locations while static assignments do not react against the time-varying factors.

\begin{figure}[ht]
\centering
\includegraphics[clip,trim=30 50 50 80,width=0.7\linewidth]{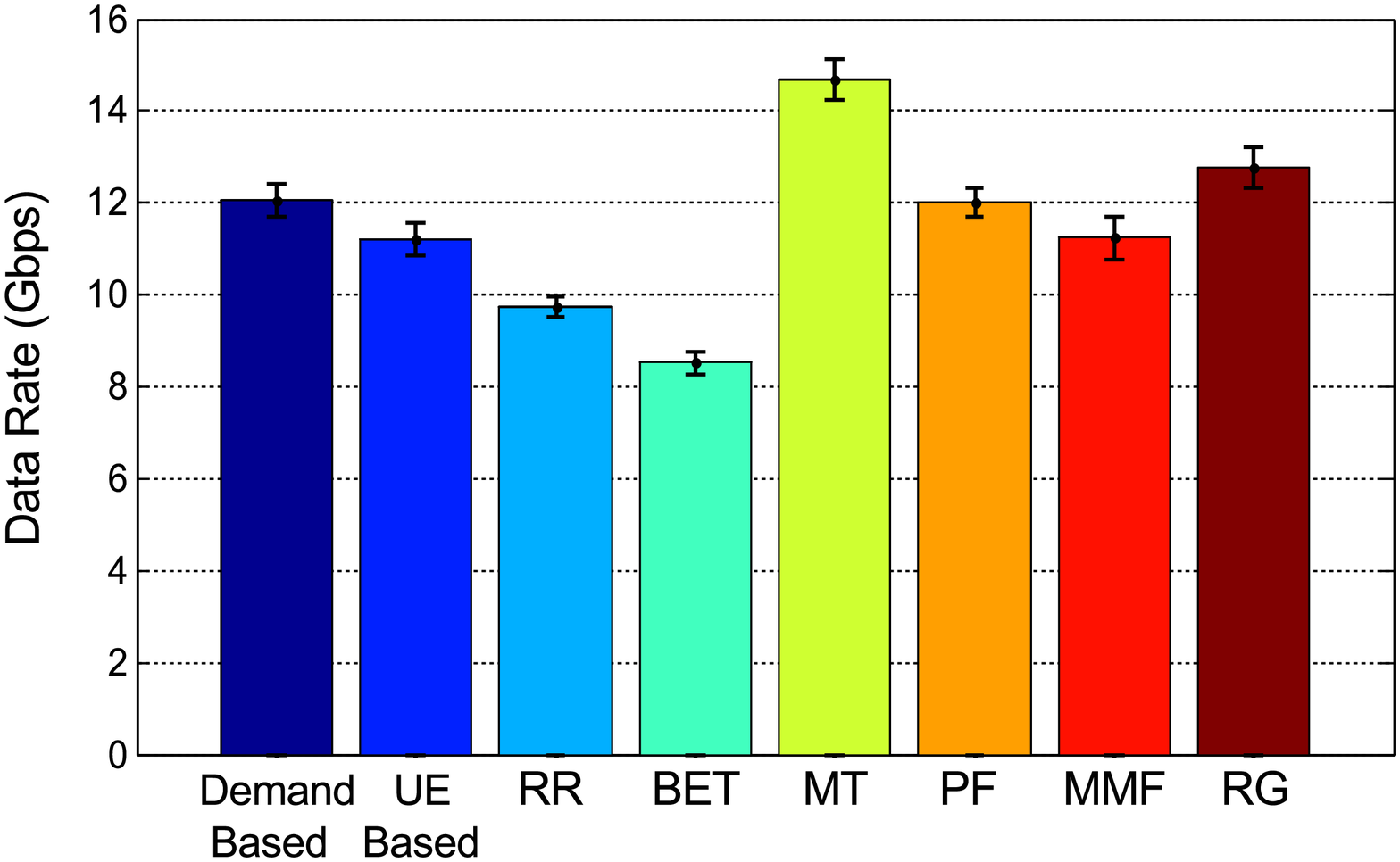}
\caption{{Data rate performances of traditional \ac{EPS} in \ac{LTE} (\textit{demand-based} and \textit{\ac{UE}-based} \ac{eNodeB} assignments) and proposed shared \ac{SDN}-based \ac{EPS} architecture with different scheduler algorithms.}}
\label{fig: Data rate performances}
\end{figure}

Data rate performances of our proposed architecture and static assignments are depicted in Fig.~\ref{fig: Data rate performances}. In contrast to fairness index, \ac{BET} gives the lowest data rate with the value of $8.52$~$\rm{Gbps}$ and this is followed by \ac{RR} with $9.73$~$\rm{Gbps}$. \ac{MT} scheduler algorithm, which gives the lowest fairness index of $0.37$, achieves the highest data rate with of $14.68$~$\rm{Gbps}$. Additionally, \ac{RG} scheduler gives $12.76$~$\rm{Gbps}$ data rate. Under the consideration that the \textit{demand-based} and \textit{\ac{UE}-based} assignments reach $12.06$~$\rm{Gbps}$ and $11.21$~$\rm{Gbps}$, our proposed architecture shows improvement with the use of \ac{MT} and \ac{RG} schedulers when comparing both static assignments as the considered schedulers try to maximize the system throughput and satisfy the demands.

\begin{figure}[ht]
\centering
\includegraphics[clip,trim=30 50 50 80,width=0.7\linewidth]{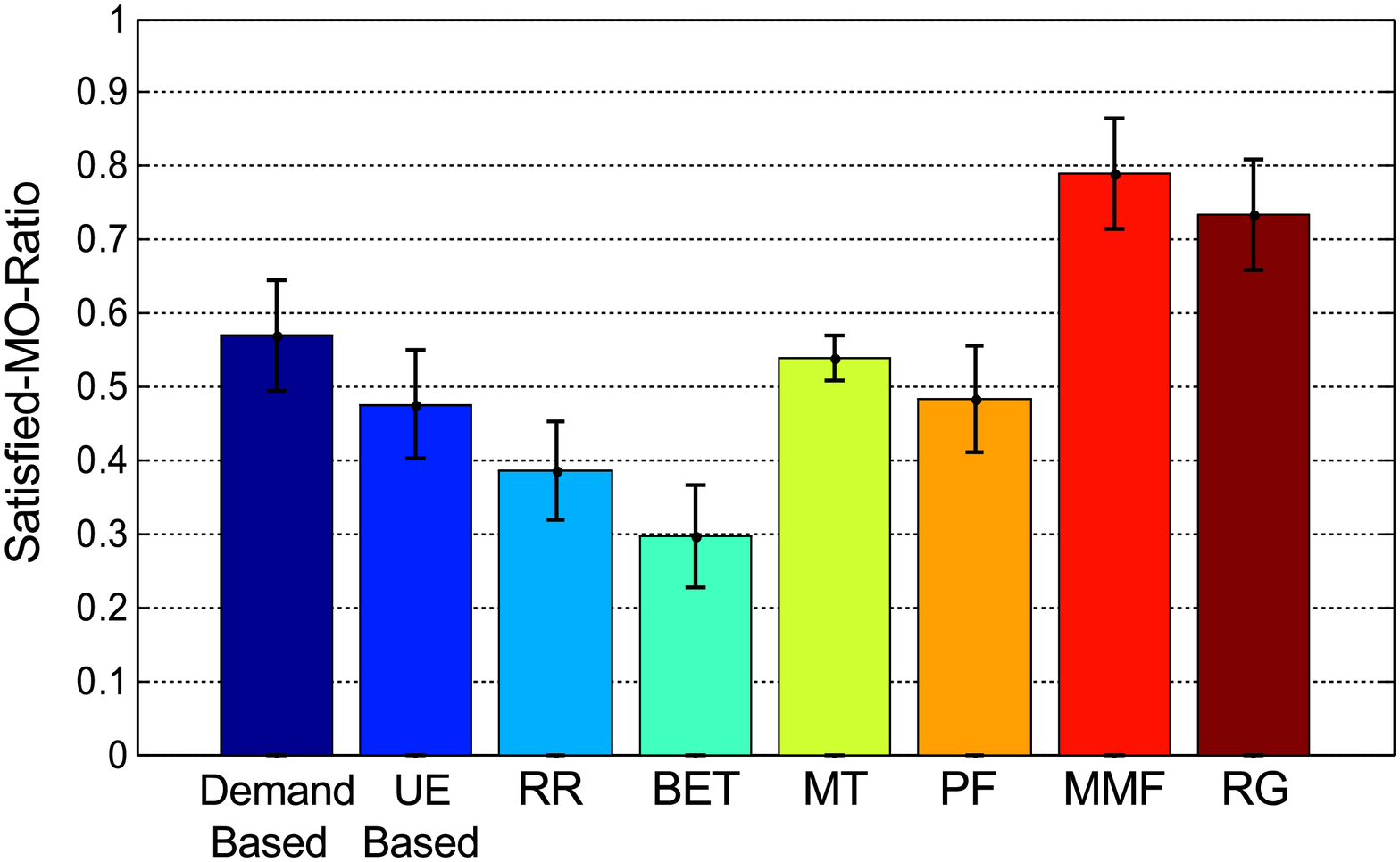}
\caption{Satisfied-MO-ratio performances of traditional \ac{EPS} in \ac{LTE} (\textit{demand-based} and \textit{\ac{UE}-based} \ac{eNodeB} assignments) and proposed shared \ac{SDN}-based \ac{EPS} architecture with different scheduler algorithms.}
\label{fig: satisfied user ratio}  
\end{figure}

In Fig.~\ref{fig: satisfied user ratio}, we show the satisfied-MO-ratio performance. Since \ac{MT} and \ac{PF} schedulers consider \ac{CSI}, \ac{MO}$-2$ can be satisfied without considering \ac{QoS} requirements, therefore, the ratios are $0.54$ and $0.48$, respectively. In \ac{MMF} and \ac{RG} with selected $\beta$ values, satisfied-user ratios are $0.79$ and $0.73$ as \textit{demand-based} and \textit{\ac{UE}-based} achieve the values of $0.61$ and $0.51$, respectively. These improvements through \ac{MMF} and \ac{RG} are due to the fact that while assigning \glspl{eNodeB} to \glspl{MO}, these allocation mechanisms consider \ac{QoS} requirements of \glspl{MO}.

Under the consideration of both fairness and satisfied-MO-ratio performances, our proposed architecture with the use of \ac{MMF} scheduler outperforms the currently deployed architecture. Similarly, \ac{RG} scheduler improves the performance when we consider both data rate and satisfied-MO-ratios. The results prove the benefits of programmable networking and centralized policy control (as in our proposed architecture) with respect to currently used traditional network architectures. The system performances in terms of fairness, data rate, satisfied-MO-ratio, both fairness and satisfied-MO-ratio and both data rate and satisfied-MO-ratios can be improved by the proposed architecture with a proper selection of scheduling algorithm such as \ac{MMF} and \ac{RG} respectively. 

\begin{figure}[ht]
\centering
\includegraphics[clip,trim= 50 0 50 0,width=0.9\linewidth]{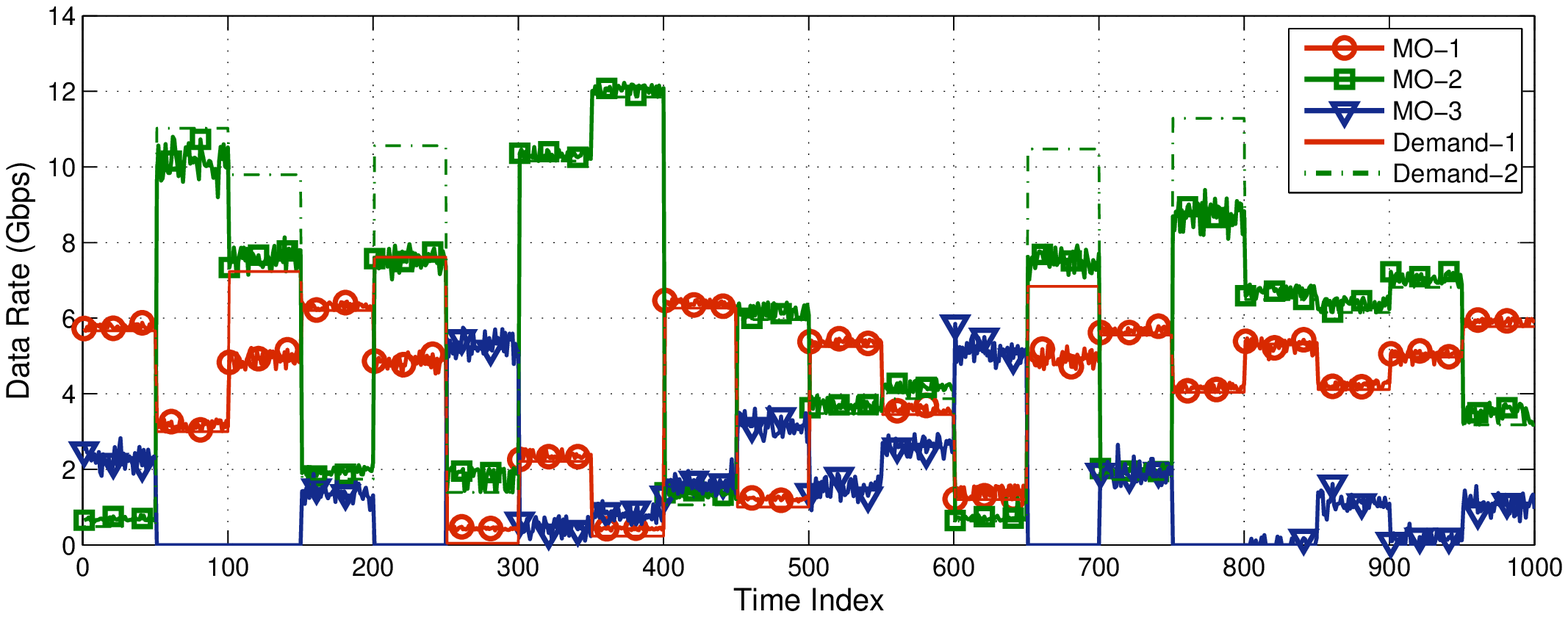}
\caption{Evolution of obtained data rates three \glspl{MO} as well as demands of two \glspl{MO} with the use of \ac{MMF} scheduler by \ac{BTP}.}
\label{allocation}
\end{figure}

In Fig.~\ref{allocation}, we present an example of \ac{eNodeB} assignment results with the use of \ac{MMF} scheduler in time varying channel and different demand conditions of each \ac{MO} corresponding to the parameters defined above. In the time interval of $300-350$, \ac{QoS}-aware \glspl{MO} which are \ac{MO}$-1$ and \ac{MO}$-2$  are satisfied and remaining \glspl{eNodeB} are assigned to \ac{MO}$-3$. Therefore, satisfied-MO-ratio reaches the highest value. On the other hand, because the demand of \ac{MO}$-2$ is lower than that of \ac{MO}$-1$ in the time interval $750-800$ and available resource is not enough to satisfy both, only \ac{MO}$-2$ achieves its demand. Consequently, \ac{MO}$-3$ is not assigned with any \glspl{eNodeB} by the \ac{BTP}. Similarly, when the time interval of $100-150$ is considered, since none of \ac{MO}$-1$ and \ac{MO}$-2$ are satisfied, \ac{MO}$-3$ cannot serve its \glspl{UE}.

\begin{figure}[ht]
\centering
\includegraphics[clip,trim=-100 350 200 0,width=0.9\textwidth]{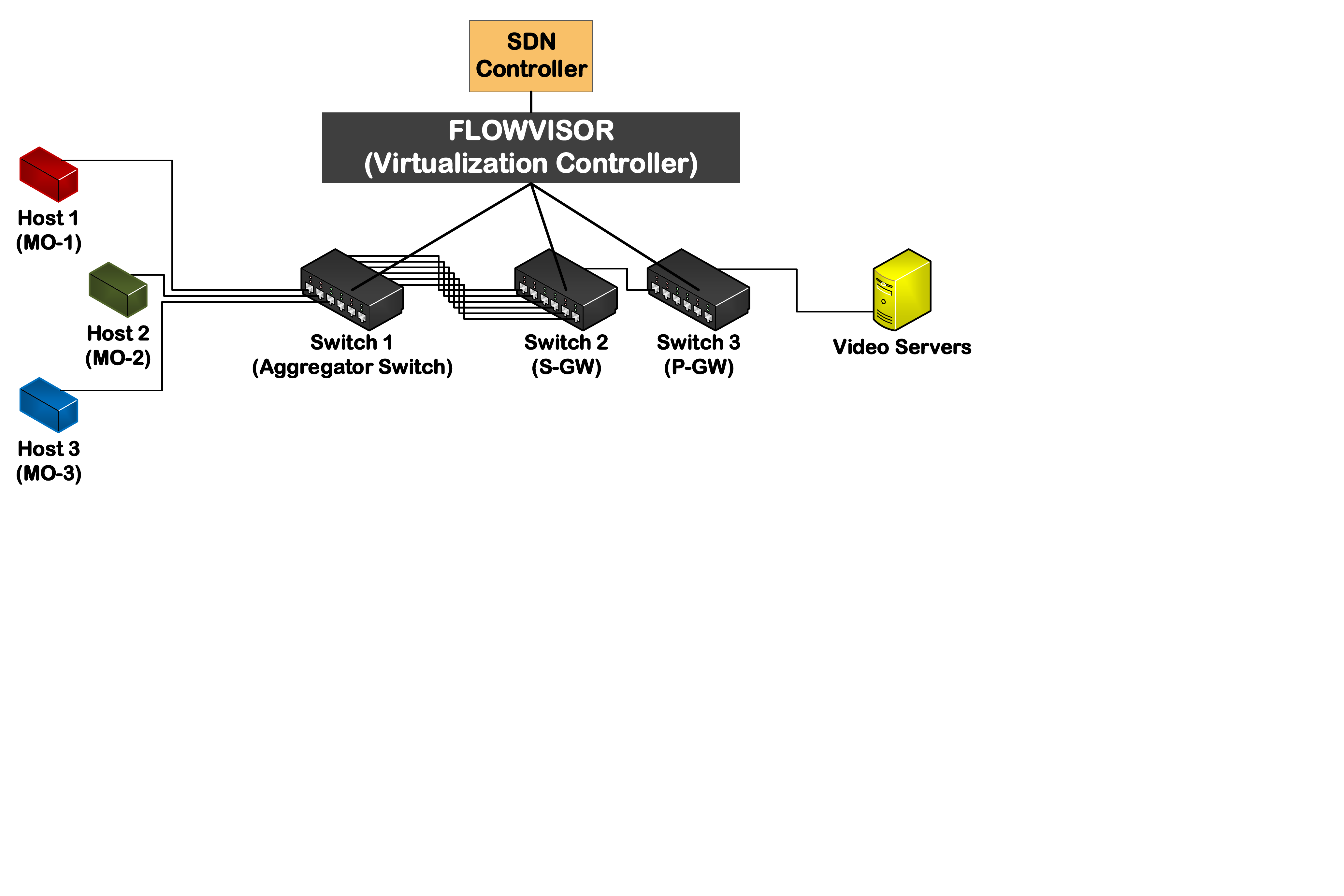}
\caption{Considered topology on Mininet.}
\label{MininetTopo}
\end{figure}

\begin{figure}[ht]
\centering
\includegraphics[clip,trim=0 50 0 50,width=\linewidth]{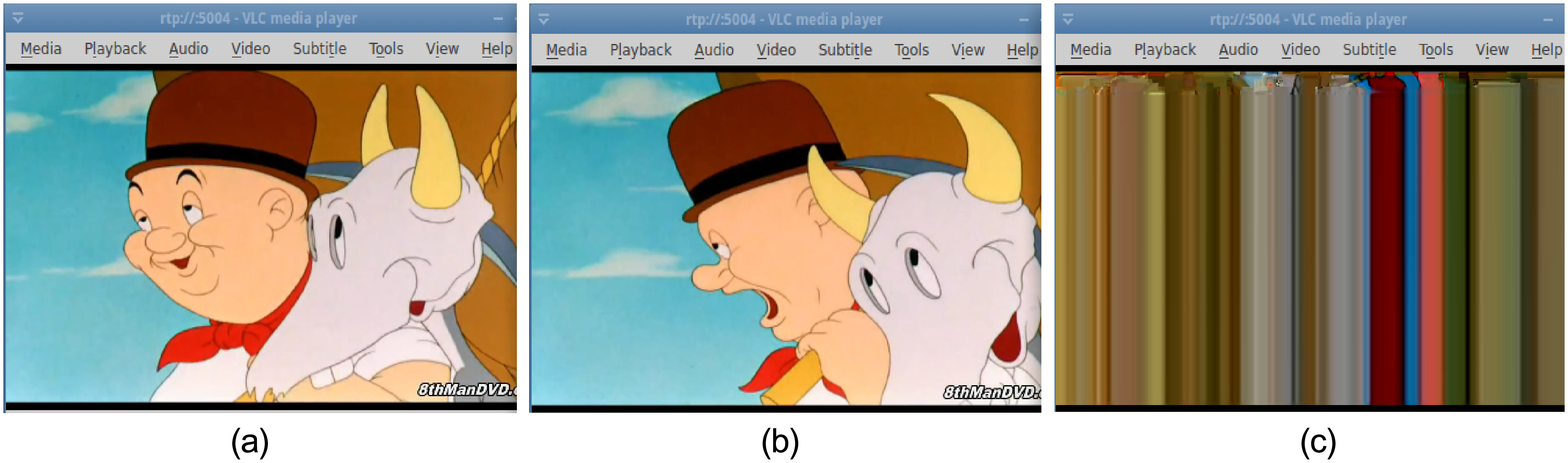}
\caption{Validations on Mininet for video quality on (a) \ac{MO}$-1$, (b) \ac{MO}$-2$ and (c) \ac{MO}$-3$.}
\label{fig:Allocation of MMF scheduler}
\end{figure}

In the second experimental results, we try to generate our proposed network architecture with the use of \ac{MMF} scheduler on Mininet, \textit{de-facto} \ac{SDN} emulator. The generated network topology is depicted in Fig.~\ref{MininetTopo} where three hosts, three OpenFlow switches, video servers and a virtualization controller exist. The hosts are assumed to be \glspl{MO} as in our proposed architecture and it is investigated whether their demands are satisfied or not after they send a request to video servers. The video data is delivered to hosts through \ac{PGW}, \ac{SGW} and aggregator switch that are virtualized through FlowVisor as in SDN-based \ac{LTE} architecture. The multiple links between aggregator switch and \ac{SGW} are likened to \glspl{eNodeB} since \glspl{eNodeB} direct the data traffic from \ac{UE} to \ac{SGW} and vice versa over \ac{GTP} tunnel in \ac{LTE} networks. After virtualization of both aggregator switch and \ac{SGW}, link assignment is performed through FlowVisor virtualization controller, similar to \ac{eNodeB} assignment. Wireshark\textsuperscript{\textregistered} is used for further analysis such as traffic monitoring and collection of the amount of transferred data. The requirement of video transfer is identified as approximately 3 Mbps. \ac{QoS}-aware users (\rm{MO}$-1$ and \rm{MO}$-2$) are assigned with links responding to their demands with the use of \ac{MMF} algorithm and the remaining links in between aggregator switch and \ac{SGW} (whose total capacity is less than $3$~$\rm{Mbps}$) are assigned to \ac{BE} user (\ac{MO}$-3$). Under the consideration of {assigned links}, video qualities are presented in Fig.~\ref{fig:Allocation of MMF scheduler}. Since \ac{QoS}-aware users are satisfied, their video quality is good enough, however, the video of \ac{BE} user is fully defective due to insufficient assignment.

\section{Conclusion and Future Work}

A detailed analysis for designing a virtualization controller which is controlled by \ac{BTP} in a shared \ac{SDN}-based \ac{EPS} architecture that can benefit both \glspl{MO} and \glspl{BTP} has been provided. The \glspl{eNodeB} become a part of resource allocation problem for \ac{BTP} as a consequence of core and backhaul network virtualization and their assignment to different \glspl{MO} are performed by \ac{BTP} based on the demands of the \glspl{MO} and the adopted scheduling algorithms. The results reveal that depending on the selection of scheduling algorithm, our proposed \ac{SDN}-based \ac{EPS} architecture outperforms the traditional \ac{EPS} with the use of both \ac{QoS}-aware (\ac{MMF} and \ac{RG}) and non-\ac{QoS}-aware (\ac{RR}, \ac{BET}, \ac{MT} and \ac{PF}) schedulers in terms of fairness, data rate and satisfied-MO-ratio, and this shows the benefits of programmable networking and centralized policy control. In addition, in order to showcase a real emulation environment of proposed architecture, the performance of \ac{MMF} scheduling algorithm is demonstrated using the Mininet emulation platform in terms of video quality for different \glspl{MO}.

Possible future extensions of this work include considering \glspl{SLA} between different \glspl{MO} when \glspl{BTP} are running them over the same infrastructure. In those scenarios, due to existence of more than two conflicting objective functions, multi-objective optimization techniques leading to Pareto optimality, instead of the simpler \ac{QoS}-based algorithms, are of interest.

\section*{Acknowledgment}
This work has been performed in the framework of the Celtic-Plus project $\rm{C2012/2-5}$ SIGMONA and T\"{U}B\.{I}TAK TEYDEB (project no 9130038) project. The organizations on the source list would like to acknowledge the contributions of their colleagues in the project.

\section*{References}
\bibliographystyle{elsarticle-num} 
%\nocite{*} 
%\bibliographystyle{IEEEtran} 
\bibliography{references}

\bigskip

\textbf{Omer Narmanlioglu} received B.Sc. and M.Sc. degrees from the Department of Electrical and Electronics Engineering at Bilkent University, Ankara, Turkey, in 2014 and Ozyegin University, Istanbul, Turkey, in 2016. He is currently working with P.I. Works and pursuing PhD degree at Ozyegin University. His research interests are physical layer aspects of communication systems and software-defined networking paradigm for cellular networks.

\textbf{Engin Zeydan} received his PhD degree in February 2011 from the Department of Electrical and Computer Engineering at Stevens Institute of Technology, USA. He has worked as an R\&D engineer for Avea, a mobile operator in Turkey, between 2011 and 2016. He is currently with T\"{u}rk Telekom Labs. His research interests are in the area of telecommunications and big data networking. 

\end{document}